\documentstyle[epsfig,longtable]{aipproc}
\newcommand{\qsq    }{\mbox{$Q^2$}}
\newcommand{\psq    }{\mbox{$P^2$}}
\newcommand{\qzm    }{\mbox{$\langle \qsq \rangle$}}
\newcommand{\gev    }{\mbox{$\rm GeV$}}
\newcommand{\gevsq  }{\mbox{$\rm GeV^2$}}
\newcommand{\mev    }{\mbox{$\rm MeV$}}
\newcommand{\der    }{\mbox{${\mathrm d}$}}
\newcommand{\ft     }{\mbox{$F_{2}^{\gamma}$}}
\newcommand{\aem    }{\mbox{$\alpha$}}
\newcommand{\aemsq  }{\mbox{$\aem^2$}}
\newcommand{\ftxq   }{\mbox{$\ft(x,\qsq)$}}
\newcommand{\flxq   }{\mbox{$\fl(x,\qsq)$}}
\newcommand{\Wvis   }{\mbox{$W_{\mathrm{vis}}$}}

\newcommand{\mc     }{\mbox{$m_{\mathrm{c}}$}}
\newcommand{\ftc    }{\mbox{$F_{2,\mathrm{c}}^{\gamma}$}}
\newcommand{\invpb  }{\mbox{$\mathrm{pb}^{-1}$}}

\newcommand{\eqt    }{\mbox{${\rm e^2_{\rm q}}$}}
\newcommand{\al     }{\mbox{$\alpha_{s}$}}
\newcommand{\wsq    }{\mbox{$W^{2}$}}
\newcommand{\fl     }{\mbox{$F_{\mathrm{L}}^{\gamma}$}}
\newcommand{\dm     }{\mbox{$\Delta\,M$}}
\newcommand{\dst    }{\mbox{${\rm D}^\star$}}
\newcommand{\etadst }{\mbox{$\eta^{{\rm D}^\star}$}}
\newcommand{\ptdst  }{\mbox{$p_{\rm T}^{{\rm D}^\star}$}}
\newcommand{\dn     }{\mbox{${\rm D}^0$}}
\newcommand{\sigcc  }{\mbox{$\sigma({\rm e}^+{\rm e}^-\to{\rm e}^+{\rm e}^-{\rm c}\bar{{\rm c}})$}}
\begin{document}
\title{First Measurement of the Photon Structure 
       Function \boldmath\ftc\unboldmath}
\author{Richard Nisius (OPAL Collaboration)%
\thanks{Invited talk given at the PHOTON 2000 Conference, Ambleside, 
        UK, August 26-31, 2000, to appear in the Proceedings.}
}
\address{CERN, CH-1211 Gen\`eve, Switzerland, Richard.Nisius@cern.ch}
%
\maketitle
%
%
\begin{abstract}
 The first measurement of \ftc\ is presented. 
 At low $x$ the measurement indicates a non-zero hadron-like component
 to \ftc.
 At large $x$ the measurement constitutes a test of perturbative QCD 
 at next-to-leading order, with only \mc\ and \al\ as free parameters,
 with a precision of ${\cal{O}}(40\%)$.
\end{abstract}
%
%
\section*{Introduction}
 For about 20 years measurements of photon structure functions give deep
 insight into the rich structure of a fundamental gauge boson, the photon.
 A recent review on this subject can be found in~\cite{NIS-9904}.
 Here, the discussion is restricted to the measurement of \ftc, which
 recently has been achieved for the first time.
 Only the main features of the analysis are given, the experimental details
 can be found in~\cite{OPALPR294}.
 \par
\begin{figure}[t]
\centerline{\epsfig{file=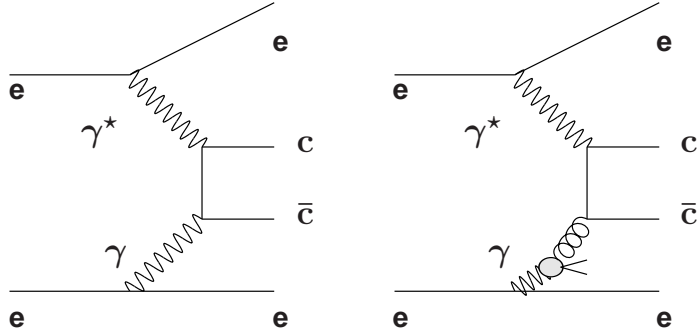,width=10cm}}
\caption{
         Examples of leading order diagrams contributing to
         (left) the point-like, and (right) the hadron-like part of \ftc.
        }\label{fig:fig01}
\end{figure}

 The differential cross-section for deep inelastic electron-photon 
 scattering, shown in Figure~\ref{fig:fig01}, is given by
%
 \begin{eqnarray}
 \frac{\der^2\sigma}{\der x\,\der\qsq} &=& \frac{2\pi\aemsq}{x\,Q^{4}}
 \left[\left(1+(1-y)^2\right) \ftxq  - y^2 \flxq \right]\, .
 \label{eqn:approx}
 \end{eqnarray}
%
 Here \qsq\ is the absolute value of the four momentum squared of the 
 exchanged virtual photon, $\gamma^\star$, $x$ and $y$ are the usual
 dimensionless variables of deep inelastic scattering and \aem\ is
 the fine structure constant.
 In experimental analyses $y^2$ is usually small.
 Consequently, the term proportional to the longitudinal structure 
 function \fl\ can be neglected and the differential cross-section is
 directly proportional to \ft, which is related to the
 sum over the quark parton distribution functions $q^{\gamma}$ of the 
 quasi-real photon, $\gamma$, via
%
\begin{eqnarray}
 \ftxq &=& x \sum_{q=u,d,s}^{c,b,t} \eqt
           \left[q^{\gamma}(x,\qsq)+\bar{q}^{\gamma}(x,\qsq)\right]\, .
 \label{eqn:F2def}
\end{eqnarray}
%
 Due to the large scale established by their masses, the contribution to
 \ft\ from heavy quarks can be calculated in perturbative QCD.
 At present collider energies only the contribution of charm quarks
 \ftc\ is important.
 Like the structure function for light quarks, \ftc\ receives contributions
 from the point-like and the hadron-like component of the photon
 shown in Figure~\ref{fig:fig01}.
 \par
%
\begin{figure}[tbh]
\centerline{\epsfig{file=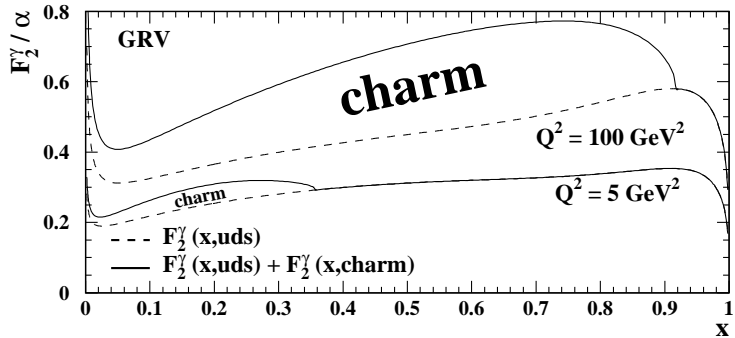,width=10cm}}
\caption{
         The structure function \ft\ for $u,d,s$ quarks alone
         and for $u,d,s,c$ quarks, as
         a function of $x$ and for two different values of \qsq. 
        }\label{fig:fig02}
\end{figure}

%
 Because of the charge of the charm quarks their contribution to \ft\
 is large and the importance increases for increasing values of \qsq, 
 as can be seen from Figure~\ref{fig:fig02}, which shows the contributions
 from light quarks and from charm quarks separately, as predicted by 
 the GRV parametrisations~\cite{GRV}.
%
%
 Charm quarks can only be produced if the photon-photon invariant mass 
 $W$ is at least twice the mass of the charm quarks \mc. 
 Using $x = \qsq/(\qsq+\wsq)$ this leads to the varying production 
 threshold in $x$ as a function of \qsq\ seen in Figure~\ref{fig:fig02}.
 \par
 Close to the production threshold, the point-like contribution to \ftc\
 is accurately approximated by the prediction of the lowest order 
 Bethe-Heitler formula.
 For quasi-real photons also the next-to-leading order
 (NLO) predictions have been calculated in~\cite{LAE-9401}.
%
\begin{figure}[tbh]
\centerline{\epsfig{file=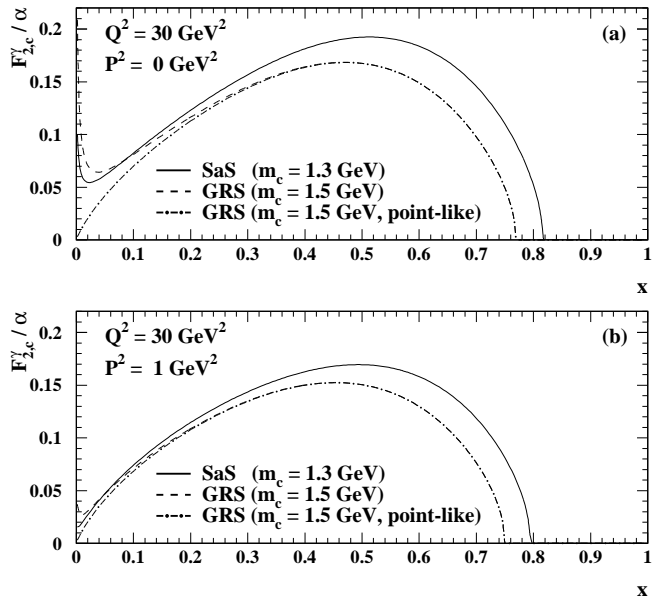,width=9.5cm}}
\caption{
         The predictions of the SaS1D (full) and the GRS
         (dash) parametrisations for $\qsq=30$~\gevsq\ and
         for (a) $\psq=0$ and (b) $\psq=1$~\gevsq.
         In addition, the point-like contribution for $\mc=1.5$~\gev 
         (dot-dash) is shown.
        }\label{fig:fig03}
\end{figure}
%
 For the hadron-like contribution the photon-quark coupling must be replaced 
 by the gluon-quark coupling, and the Bethe-Heitler formula has
 to be integrated over the allowed range in fractional momentum of the gluon 
 using a parametrisation of the gluon distribution function of the photon,
 see e.g.~\cite{NIS-9904}.
 \par
 The predicted behaviour of the point-like and hadron-like component
 of \ftc\ for different values of \mc\ is shown in Figure~\ref{fig:fig03},
 using the SaS1D~\cite{SAS} and GRS~\cite{GLU-9501} parametrisations.
 For $x>0.1$ the structure function is saturated by the point-like
 component which is only slowly suppressed for increasing
 virtualities \psq\ of the quasi-real photon.
 In contrast, the hadron-like contribution dominates at small values of
 $x$ and decreases much faster for increasing \psq.
 Finally, lowering the mass of the charm quarks leads to a higher 
 threshold in $x$.
 \par
 Given this predicted behaviour, the region of $x>0.1$ can be used to
 test a purely perturbative NLO QCD prediction with only 
 \mc\ and \al\ as free parameters, and the low $x$ behaviour 
 mainly probes the gluon distribution function of the photon.
%
%
\section*{Monte Carlo Models}
 The LO Monte Carlo generators HERWIG 5.9~\cite{MAR-9201} and
 Vermaseren~\cite{Vermas} are used, both with $\mc=1.5$~\gev.
 In HERWIG charm production is modelled using matrix elements for massless
 charm quarks, together with the GRV parametrisation 
 for the parton distributions of the photon, again for massless charm quarks.
 The effect of the charm quark mass is only accounted for by not
 simulating events with $W<2\mc$.
 Due to the massless approach used in HERWIG and the crude treatment at 
 threshold, the
 predicted charm production cross-section is likely to be too large.
 The Vermaseren generator is based on the Quark Parton Model (QPM)
 and consequently does not take into account the hadron-like component
 of the photon structure.
 However, the complete dependence of the cross-section on the
 different photon helicities is modelled. 
%
%
\section*{The measurement of \boldmath\ftc\unboldmath}
 The measurement of \ftc\ proceeds along the same lines as the measurement
 of \ft\ with the addition of the identification of the charm quarks via
 the reconstruction of \dst\ mesons.
 \par
 Events are selected with an energy of the scattered electron
 above 50~\gev, measured in the angular ranges (a) $33 - 55$~mrad 
 or (b) $60 - 120$~mrad from either beam direction, thereby covering 
 the approximate range in \qsq\ of $5-100$~\gevsq.
 The visible hadronic mass \Wvis\ is required to be below 60~\gev.
 Charm quarks are identified via $\dst\rightarrow\dn\pi$, followed by
 $\dn\rightarrow K\pi$ or $\dn\rightarrow K\pi\pi\pi$.
 Using $f({\rm c}\rightarrow D^\star) = 0.235\pm0.011$ and the branching
 ratios of the \dn\ decay modes of $0.02630\pm0.00082$ and
 $0.0519\pm0.0029$, this analysis covers only about 4$\%$ of all 
 events containing a pair of charm quarks. 
 For a clear acceptance the \dst\ mesons are further required to fulfill,
 $\vert\etadst\vert<1.5$ and $\ptdst>1 \mbox{ or }3$~\gev\ for (a) or (b).
 Together with a typical selection efficiency of about 25$\%$ only
 about 1$\%$ of all ${\rm c}\bar{\rm c}$ events are positively identified.
 \par
%
\begin{figure}[tbh]
\centerline{\epsfig{file=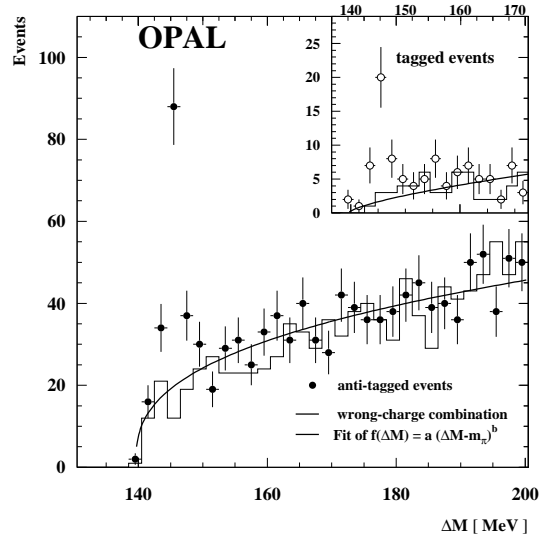,width=7cm}}
\caption{
         Mass difference \dm\ for the anti-tagged and tagged sample.
        }\label{fig:fig04}
\end{figure}
%
 Figure~\ref{fig:fig04} shows the distribution of the difference between
 the \dst\ and the \dn\ candidate mass.
 In both samples a clear peak is visible around $\dm = 145.4$~\mev.
 Subtracting the background, obtained from a fit to the upper sideband of 
 the signal, $29.8\pm 5.9({\rm stat})$ \dst\ mesons are found in the peak
 region.
 \par
%
\begin{figure}[tbh]
\centerline{\epsfig{file=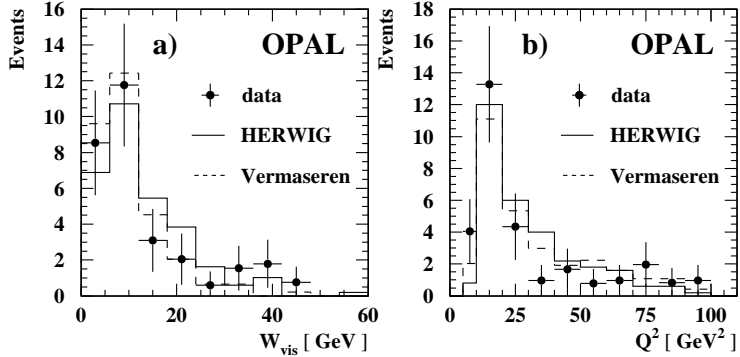,width=10cm}}
\caption{
         The distributions of a) the visible invariant mass \Wvis,
         and b) the negative four-momentum squared \qsq.
        }\label{fig:fig05}
\end{figure}
%
 Figure~\ref{fig:fig05} shows the distributions of \Wvis\ and
 of the measured \qsq\ in comparison to the predictions of the 
 HERWIG and Vermaseren Monte Carlo generators normalised to the 
 number of data events. 
 Both Monte Carlo generators give a good description of the shape of the
 data distributions.
 \par
 The cross-section for \dst\ production is determined in the 
 well-measured kinematic range described above.
 Based on this and the extrapolation factors obtained from the Monte
 Carlo models the full cross-section \sigcc\ and \ftc\ are evaluated
 in two bins of $x$ with $0.0014<x<0.1$ and $0.1<x<0.87$.
 \par
 For $x>0.1$ the predictions of the Vermaseren and HERWIG 
 Monte Carlo models are very similar. 
 In contrast, for $x<0.1$, there are two main differences. 
 Firstly the selection efficiency for events with \dst\ mesons 
 fulfilling the kinematical requirements is different, and secondly,
 and even more important, the predicted cross-section within the invisible 
 phase-space is largely different for the two models, resulting in 
 different extrapolation factors, 12.9/5.1 for HERWIG/Vermaseren.
 \par
 Since the hadron-like contribution is neglected in the QPM,
 the Vermaseren cross-section is much smaller than the 
 LO and the NLO cross-section for $x<0.1$.
 In contrast, mainly due to the massless approach taken, the 
 prediction from HERWIG is higher than the cross-section
 from the LO and the NLO calculation.
 Therefore it is likely that the correct cross-section,
 and therefore the correct extrapolation factor, lies within 
 the range of the two Monte Carlo predictions.
 \par
\begin{figure}[tbh]
\centerline{\epsfig{file=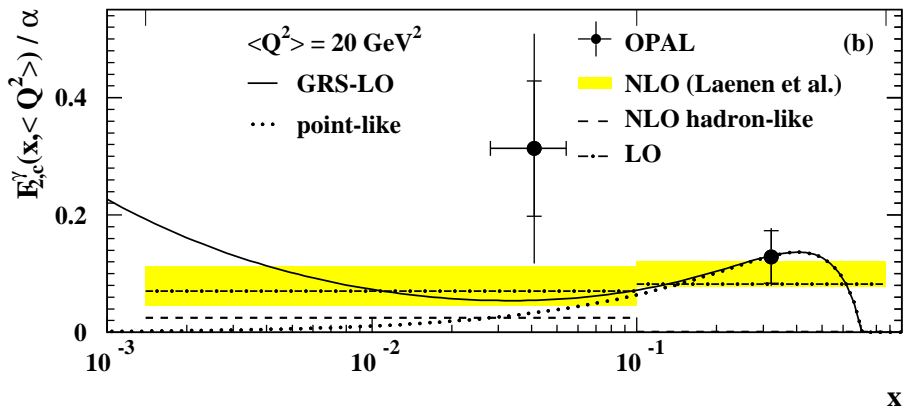,width=12cm}}
\caption{
         \ftc\ compared to several predictions explained in the text. 
        }\label{fig:fig06}
\end{figure}

 The measured \ftc\ for $\qzm= 20$~\gevsq\ is shown in Figure~\ref{fig:fig06}.
 The central values are obtained by averaging the results using the 
 HERWIG and Vermaseren Monte Carlo models, and half the difference is
 taken as extrapolation error, which dominates the uncertainty for $x<0.1$.
 The NLO prediction is based on $\mc=1.5$~\gev,
 the renormalisation and factorisation scales are chosen to
 be $\mu_{\rm R} = \mu_{\rm F} = Q$, and the hadron-like contribution
 to \ftc\ uses the GRV parametrisation.
 The NLO corrections are small for the whole $x$ range.
 The band for the NLO calculation is evaluated by varying \mc\ between 1.3
 and 1.7~GeV and using $Q/2 \le \mu_{\rm R}=\mu_{\rm F}\le 2~Q$.
 \par
%
%
 For $x>0.1$
 the error of the measured cross-section is dominated by the statistical
 uncertainty, and the NLO calculation with only \mc\ and \al\ as
 free parameters is in good agreement with the data.
 In contrast, for $x<0.1$, the result suffers from the strong model 
 dependence discussed above.
 Despite this uncertainty the corrected data suggest a 
 cross-section which is above the purely point-like component, i.e.
 the hadron-like component of \ftc\ is non-zero.
%
%
\subsection*{Conclusion and Outlook}
 In conclusion, for $x>0.1$, the purely perturbative NLO calculation is
 in good agreement with the measurement and for $x<0.1$, 
 the measurement suggests a non-zero hadron-like component of \ftc.
 \par
 By using the massive matrix elements available in HERWIG6.1 and the full
 integrated luminosity of more than 500~\invpb\ of the LEP2 programme, the
 measurement is likely to be improved considerably, both concerning
 the statistical and the systematic error.
 \newline
%
%
 \mbox{  }\\
 {\bf Acknowledgement:}\\
 I wish to thank the organisers of this interesting conference, and
 especially Alex Finch. They created a fruitful atmosphere throughout
 the meeting and made attending the meeting a very nice experience. 
%
%

\end{document}